\documentclass[10pt]{article}
\usepackage{epsf}
\usepackage{proofenv}
\usepackage{times}

\newcommand\cG{{\cal G}}
\newcommand\cV{{\cal V}}
\newcommand\cP{{\cal P}}

\newcommand\mb[1]{\mbox{{\scriptsize #1}}}

\newcommand\genmachine[1]{\mbox{${\cal M}_{g}(#1)$}}
\newcommand\recmachine[1]{\mbox{${\cal M}_{r}(#1)$}}
\newcommand\tapes[7]{\mbox{{\scriptsize \begin{math}\begin{array}{r|c|l}\multicolumn{3}{l}{} \\
 \cline{2-2} & \mb{#1} & \mb{#2} \\ 
 \mb{#3}     & \mb{#4}     & \mb{#5} \\ 
 \mb{#6}     & \mb{#7}     & \\ \cline{2-2}
 \multicolumn{3}{l}{}\end{array}\end{math}}}}
\newcommand\linedtapes[7]{\tapes{\makebox[2em]{#1}}{#2}{#3}{\makebox[2em]{#4}}{#5}{#6}{\makebox[2em]{#7}}}
\newcommand\nada{\mbox{$\emptyset$}}
\newcommand\productionsym{\longrightarrow}
\newcommand\production[2]{\mbox{#1 $\productionsym$ #2}}
\newcommand\trans[1]{\stackrel{*}{#1}}
\newcommand\move{\longmapsto}
\newcommand\nmove[1]{\stackrel{#1}{\longmapsto}}
\newcommand\cmove{\stackrel{\mbox{c}}{\longmapsto}}
\newcommand\reduction[2]{\mbox{$#1 \longrightarrow #2$}}

\newtheorem{definition}{Definition}[section]
\newtheorem{lemma}[definition]{Lemma}
\newtheorem{proposition}[definition]{Proposition}
\newtheorem{corollary}[definition]{Corollary}

\begin{document}

\title{An Analysis of Lambek's Production Machines\thanks{This paper
is essentially the same as one that appeared in \emph{RAIRO Informatique Th\'eorique et Applications}, 31(5), pp. 483--497, 1997.}}
\author{Riccardo R. Pucella\\[.1in]Bell Laboratories\\Lucent Technologies}
\date{}

\maketitle

\begin{abstract}
Lambek's production machines may be used to generate and recognize
sentences in a subset of the language described by a production
grammar. We determine in this paper the subset of the language of a
grammar generated and recognized by such machines.
\end{abstract}

\section{Introduction}

The focus of this paper is the mechanical generation and recognition
of sentences from a production grammar \cite{Hopcroft69,Lambek89},
which are known in mathematics as semi-Thue systems and in linguistics
as rewriting systems or generative grammars. The latter, linguistics,
is an important area of 
application for production grammars. They were used to study French
and Latin conjugation \cite{Lambek75,Lambek79} and kinship terminology
in English \cite{Lambek86} and other languages
\cite{Lambek81,Bhargava83,Bhargava92,Bhargava95}.
Production grammars were also provided for subsets of English and
French \cite{Lambek93a,Pucella96a} and used in a naive approach to
syntactic translation \cite{Pucella96a}.

To generate and recognize sentences in languages defined by a
production grammar, Lambek combined two pushdown automata into a
single machine \cite{Lambek93} and gave examples of the execution of
the machine on simple sentences taken from a grammar describing a
subset of English. 

Our previous work \cite{Pucella96a} indicates that Lambek's production
machines generate and recognize a subset of the language of a grammar
--- in other words, they do not generate or recognize sentences not in
the language. This paper analyzes the machines in order to determine
exactly which subsets of the language are generated and
recognized. The sublanguage generated is generally a proper subset of
the language, which we call the leftmost language. Correspondingly, the
sublanguage recognized, also generally a proper subset of the
language, may be seen as a dual to the leftmost language.

\section{Production grammars}

We review in this section the fundamental material needed in the
paper. We assume the reader is acquainted with the theory of formal
languages, so that only a short overview of the notation is necessary.

A production grammar is a tuple $\cG=(\cV,\cV_{i},\cV_{t},\cP)$ where
$\cV$ (the vocabulary or alphabet) is a finite set, $\cV_{i}$ and $\cV_{t}$
(the initial and terminal vocabularies) are subsets of $\cV$, and $\cP$
(the productions) is a finite or at least recursive set of pairs
$(\Gamma,\Delta)$ with $\Gamma$ and $\Delta$ strings of elements of
$\cV$. We usually represent an element $(\Gamma,\Delta)$ of $\cP$ as
\production{$\Gamma$}{$\Delta$}. An element of $\cV_{t}$ is called a
\emph{terminal} symbol, while an element of $\cV-\cV_{t}$ is called a
\emph{nonterminal} symbol. A string of elements of $\cV$ will typically
be denoted by a greek letter, and individual elements of $\cV$ by
capital roman letter. 

From any production grammar $\cG=(\cV,\cV_i,\cV_t,\cP)$ one obtains
the \emph{dual} grammar of $\cG$ by taking
$\cG^{-1}=(\cV,\cV_t,\cV_i,\cP^{-1})$ where 
$\cP^{-1}$ is the set of all pairs $(\Delta,\Gamma)$ such that
$(\Gamma,\Delta) \in \cP$. 

A production \production{$\Gamma$}{$\Delta$} is \emph{applicable} to a
string $\sigma$ of element of $\cV$ if $\sigma$ is of the form
$\sigma_{1}\Gamma\sigma_{2}$. The \emph{application} of
\production{$\Gamma$}{$\Delta$} to $\sigma$ is the string
$\sigma_{1}\Delta\sigma_{2}$. A production \production{$\Gamma$}{$\Delta$}
is \emph{leftmost applicable} to a string $\sigma$ if $\sigma$ is of
the form $\sigma_{1}\Gamma\sigma_{2}$ and for any production
\production{$\Gamma'$}{$\Delta'$}, if $\sigma$ is of the
form $\gamma_{1}\Gamma'\gamma_{2}$, then $|\Gamma|\leq|\Gamma'|$ and
$|\sigma_{1}| \leq |\gamma_{1}|$. 

We define the leftmost reduction relation on strings of elements of
$\cV$ as follows: let $\sigma_{1}\longrightarrow\sigma_{2}$ if a
production of $\cG$ is leftmost applicable to $\sigma_{1}$ and
$\sigma_{2}$ is the application of the production to $\sigma_{1}$. A
\emph{sentence} is a string of terminal symbols in $\cV_{t}$. The
\emph{leftmost language} of a grammar $\cG$ is the set of all
sentences that 
can be derived via $\trans{\longrightarrow}$ starting from symbols in
$\cV_{i}$. If we define a reduction relation using the notion of
applicability instead of leftmost applicability, the set of sentences
that can be derived is called the \emph{language} of the grammar. For
emphasis, we sometimes refer to the language as the \emph{full
language} of the grammar. It is clear that the leftmost language of a
grammar is a subset of the full language. The following grammar shows
that the inclusion may be proper:
\begin{eqnarray*}
\mbox{S} & \productionsym & \mbox{ABC} \\
\mbox{AB} & \productionsym & \mbox{x} \\
\mbox{BC} & \productionsym & \mbox{y} \\
\mbox{C} & \productionsym & \mbox{z} \\
\mbox{A} & \productionsym & \mbox{w}
\end{eqnarray*}
The full language of this grammar is $\{ \mbox{xz},\mbox{wy} \}$, and
the leftmost language is $\{ \mbox{wy} \}$.

We assume in this paper that all grammars under consideration are
\emph{well-formed}, in the sense that all reduction sequences
ultimately lead to 
sentences --- string of terminal symbols. This among other things implies
that there is at least one production for each initial symbol in $\cV_{i}$.
We shall also assume, as it is usually done, that there is no empty
production and that no terminal appears on the left side of a production. 

Let us now present three transformations one needs to perform on a
grammar $\cG$ to make it suitable for treatment by the machine we
introduce in the next  section. A requirement of the transformations
is that they preserve the leftmost language of the
untransformed grammar. 

The first transformation takes a grammar $\cG$ with initial vocabulary
$\cV_{i}$ 
and produces a new grammar $\cG'$ with a unique initial symbol, say
S (this symbol must be a new symbol not originally in $\cV$). The
transformation simply consists of adding a new production for every initial
symbol of $\cG$. For example, if $\cV_{i}=\{$A$,$B$,$C$\}$, we add the
productions 
\begin{eqnarray*}
\mbox{S} & \productionsym & \mbox{A} \\
\mbox{S} & \productionsym & \mbox{B} \\
\mbox{S} & \productionsym & \mbox{C}
\end{eqnarray*}
and let the new initial vocabulary be $\cV_{i}=\{$S$\}$. It is clear that the
leftmost language of $\cG$ is preserved by this
transformation. The second transformation is the process of normalization.
A production \production{$\Gamma$}{$\Delta$} is called \emph{normal} if
both $\Gamma$ and $\Delta$ have length 1 or 2. A normal grammar is a grammar
in which every production is normal. Normalization produces a normal grammar
from a grammar, while preserving the leftmost language of the grammar. The
transformation consists in iterating the following production replacements
(the symbol N is always taken to be a new symbol not in $\cV$ at every
production replacement): 
\begin{eqnarray*}
\production{$\Gamma$}{AB$\Delta$} & \Rightarrow & 
 \production{$\Gamma$}{N$\Delta$} \\
 & & \production{N}{AB} \\
\production{AB$\Gamma$}{$\Delta$} & \Rightarrow & 
 \production{N$\Gamma$}{$\Delta$} \\
 & & \production{AB}{N} 
\end{eqnarray*}
For the last production replacement, the same symbol N must be used for
all productions with the same left side, e.g. AB$\Gamma$. To see why the
leftmost language of the original grammar is preserved, consider the
two cases that arise: if \production{$\Gamma$}{AB$\Delta$} is leftmost
applicable, so is \production{$\Gamma$}{N$\Delta$}, and once applied, by
leftmost reduction and since no other production may involve the newly
introduced symbol N, the next production to apply must be
\production{N}{AB}; similarly, if \production{AB$\Gamma$}{$\Delta$} is leftmost
applicable, so is \production{AB}{N}, and once applied, the leftmost
applicable productions include \production{N$\Gamma$}{$\Delta$} (again, since
the newly introduced symbol N cannot appear in other productions not of
the form \production{N$\Gamma$}{...}).

The next transformation we consider isolates the generation of terminal
symbols into their own production. Assuming the grammar under consideration
is normal, iterate the following productions replacement (the symbols
N,N$_{1}$,N$_{2}$ are taken to be new symbols not in $\cV$ for every
replacement, and the symbols t,t$_{1}$,t$_{2}$ are taken to be terminal
symbols): 
\begin{eqnarray*}
\production{$\Gamma$}{At} & \Rightarrow & \production{$\Gamma$}{AN} \\
 & & \production{N}{t} \\
\production{$\Gamma$}{tA} & \Rightarrow & \production{$\Gamma$}{NA} \\
 & & \production{N}{t} \\
\production{$\Gamma$}{t$_{1}$t$_{2}$} & \Rightarrow &
 \production{$\Gamma$}{N$_{1}$N$_{2}$} \\
 & & \production{N$_{1}$}{t$_{1}$} \\
 & & \production{N$_{2}$}{t$_{2}$} 
\end{eqnarray*}
It is clear that this transformation preserves the leftmost language of the
original grammar.

Please note that the first transformation applied to a grammar $\cG$
has the same 
effect as the last transformation when one considers the dual grammar
$\cG^{-1}$, namely to isolate the production of the (then terminal)
symbol S.

The last transformation has the following interesting (and useful)
consequence: 
\begin{lemma}
Given $\cG$ a grammar to which the last transformation above has been
applied. If a terminal symbol is produced after leftmost applications
of productions, then every symbol to the left of that terminal symbol
will also be a terminal symbol.
\end{lemma}
\begin{proof}
By the last transformation applied to the given grammar, since a
terminal is produced, then the leftmost applicable production must
have been of the form \production{N}{t} with t the produced terminal
symbol. Assume that there are nonterminals to the left of that
terminal. Since no new nonterminal has been introduced, no terminal
may be used on the left of a production, and the grammar is assumed to
be well-formed, there must exist a production applicable to
nonterminals on the left of the terminal. But this contradicts the
fact that the production \production{N}{t} was leftmost.
\end{proof}

\begin{figure}[t]
\centerline {\mbox{\epsfxsize=307.5pt\epsfysize=187.5pt\epsffile{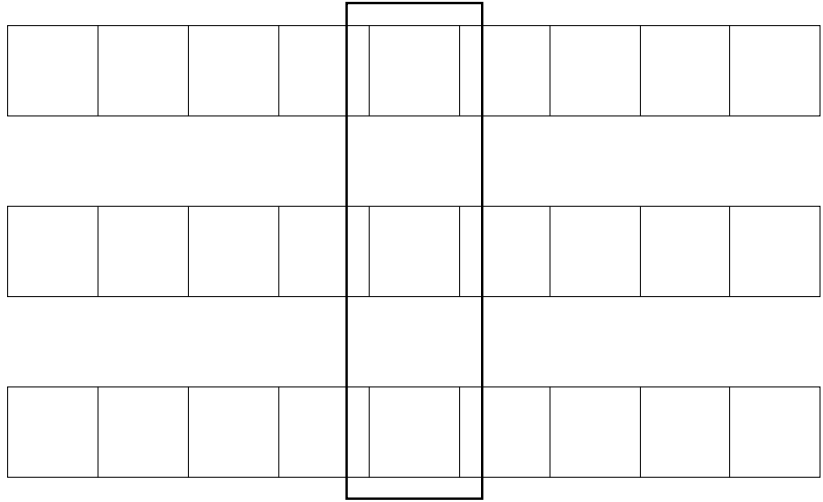}}}
\caption{Production machine}
\label{fig:one}
\end{figure}

\section{Production machines}

Lambek describes in \cite{Lambek93} a machine that allows us to
generate and recognize sentences from a production grammar. A
\emph{production machine} \cite{Lambek93,Lambek93a} corresponds
roughly to 
a combinaison of two pushdown automata. It consists of three
potentially infinite tapes  subdivided into squares. The middle tape
is the input/ouput tape, the top and bottom tapes are storage
tapes. Only one square in each taped is \emph{scanned} at any given
point in time. The two storage tapes can move in either direction,
whereas the input/output tape moves only from right to left. The tapes
are positioned so that all three scanned squares are aligned (see
Figure \ref{fig:one}). 

Seven moves are defined for production machines, parametrized by a given
grammar $\cG$. The moves involve the scanned squares of the tapes:

\begin{eqnarray*}
\linedtapes{C}{}{}{\nada}{}{}{(A)} & \nmove{1} &
 \linedtapes{\nada}{}{}{C}{}{}{(A)} \\
\linedtapes{\nada}{}{}{B}{}{}{\nada} & \nmove{2} &
 \linedtapes{left}{}{}{\nada}{}{}{B} \\
\linedtapes{C}{}{}{B}{}{}{(A)} & \nmove{3} &
 \linedtapes{right}{}{}{stay}{}{}{stay} \\
\linedtapes{\nada}{}{}{B}{}{}{A} & \nmove{4} &
 \linedtapes{stay}{}{}{stay}{}{}{left} \\
\linedtapes{\nada}{}{}{B}{}{}{(A)} & \nmove{5} &
 \linedtapes{(D)}{}{}{C}{}{}{right} \mbox{if \production{(A)B}{C(D)} is in
 $\cP$} \\
\linedtapes{\nada}{}{}{D}{}{}{\nada} & \nmove{6} &
 \linedtapes{left}{}{}{left}{}{}{stay} \mbox{if D $\in \cV_{t}$} \\
\linedtapes{\nada}{}{}{\nada}{}{}{(A)} & \nmove{7}  &
 \linedtapes{stay}{}{}{left}{}{}{stay}
\end{eqnarray*}

The $(\cdot)$ notation indicates that the scanned square may or may not be
empty, and $\nada$ represents an empty square. A mention of ``left'',
``right'', ``stay'' means that the corresponding tape should be
moved left, right or stay in the current position. We use the
expression ``move $\nmove{5}$ via production $P$'' to explicitely state
which production is involved in the move.

The machine may be used either to generate sentences from the grammar
or to recognize 
sentences in the grammar. Those two activities involve different
subsets of the general moves presented above, and different starting
and ending states for the machine. We will therefore speak
of production machines as though there were two types of machines: the
generative machine \genmachine{\cG} corresponding to a grammar $\cG$ and the
recognitive machine \recmachine{\cG} corresponding to a grammar $\cG$.

The generative machine of $\cG$ has the following initial and terminal states:
\[ \mbox{Initial:} \tapes{S}{}{}{\nada}{}{}{\nada} \hspace{.4in}
\mbox{Terminal:} \tapes{\nada}{}{$<$sentence$>$}{\nada}{}{}{\nada} \]
The machine is defined with respect to the grammar $\cG$, and the moves
that should be attempted in order are the following: 5, 6, 1, 2, 
3, 4. We say that a sentence $\sigma$ is \emph{producible} by
\genmachine{\cG} 
if the machine starts in the initial state and ends up in a state 
\[ \tapes{\nada}{}{$\sigma$}{\nada}{}{}{\nada} \]

The recognitive machine of $\cG$ has the following initial and terminal
states:
\[ \mbox{Initial:} \tapes{\nada}{}{}{\nada}{$<$sentence$>$}{}{\nada}
\hspace{.4in} \mbox{Terminal:} \tapes{\nada}{}{}{\nada}{}{}{S} \] 
The machine is defined with respect to the dual grammar $\cG^{-1}$ and
the moves that should be attempted in order are the following: 5, 7,
1, 2, 3, 4. We say that a sentence $\sigma$ is \emph{recognizable} by 
\recmachine{\cG} if it ends in the terminal state after starting in a state
\[ \tapes{\nada}{}{}{\nada}{$\sigma$}{}{\nada} \]

We refer the reader to \cite{Lambek93} for sample executions of the
machine to generate and recognize sentences in a simple grammar for the
English language.

One look at the moves of a production machine shows that the machine is
fundamentally nondeterministic. Indeed, move $\nmove{5}$ is used in a
nondeterministic 
way if more than one production with a left side of (A)B is present in the
grammar. For a generative production machine, this allows the machine to
generate different sentences. For a recognitive machine, this
introduces a complexity: possibly only one nondeterministic choice of
production to apply next leads to the terminating state of the machine, as
some examples in \cite{Lambek93} show. Hence, a recognitive production
machine must consider concurrently all the possible applications of move 5
and terminate when one leads to the terminating state. A sentence
$\sigma$ is therefore recognizable if one of the concurrent
consideration of an application of move $\nmove{5}$ of the recognitive
production machine reaches the terminal state.

\section{Generation}

We analyze in this section the generative production machine
\genmachine{\cG} of a given grammar $\cG$. We show that the language
generated by \genmachine{\cG} is exactly the leftmost language of $\cG$: a
sentence $\sigma$ is producible by \genmachine{\cG} if and only if
$\sigma$ is in the leftmost language of $\cG$. Without loss of
generality, we may assume that the grammar $\cG$ under consideration is
a normal grammar with a unique initial symbol S and with a
unique production corresponding to the generation of every terminal
symbol. As we saw earlier, any grammar may be transformed into such a
grammar defining the same leftmost language.

The idea underlying the proof is straightforward. Given a
grammar $\cG$ and a generative production machine \genmachine{\cG}, we
show that the graph corresponding to the leftmost reduction relation
is isomorphic to a graph corresponding to the moves of the
machines. Therefore, a string in the leftmost language of $\cG$ obtained
by leftmost reductions may be generated by the machine following the
moves specified by the isomorphism, and vice-versa.

The main operational tool we use is a transition graph. Given a set
$D$, a subset $I$ of $D$ and a non-transitive relation $<$ over $D$,
define a family of subset of $D$ by the equations
\begin{eqnarray*}
S_{0} & = & I \\
S_{n+1} & = & \{ b : a < b \mbox{ for some $a\in S_{n}$} \} 
\end{eqnarray*}
The transition graph of $<$ generated by $I$ is the graph with nodes
in $\cup_{n=0}^{\infty}S_{n}$ and an edge between $a,b\in
\cup_{n=0}^{\infty}S_{n}$ if and only if $a<b$. Define a layer of the
transition graph $T$ over $<$ generated by $I$ to be the set of all element
of the graph at a certain distance of an element of the initial subset,
$lay_{i}(T) = \{ a : \exists a_{0},\ldots,a_{i-1}\in T \mbox{ such
that }
a_{0}\in I \mbox{ and } a_{0}<\cdots < a_{i-1} < a \}$. If $T$ is defined by
the above equations for $S_{0}$ and $S_{n+1}$, it is not hard to see that
$lay_{i}(T) = S_{i}$.

For a given grammar $\cG$ with initial symbol S, the leftmost reduction
relation over strings in $\cV^{*}$ lead to the transition graph of
\reduction{}{} generated by $\{$S$\}$, which we will denote by ${\cal L}$. It
is this transition graph that we will show is isomorphic to a transition
graph derived from the moves of the generative machine.

Taking the $\move$ relation over the states of the machine also leads to a
transition graph, but it is easily seen to be much larger than the
transition graph ${\cal L}$, since for every production application (which
corresponds to a move $\nmove{5}$), there are other administrative moves
that the machine needs to perform. However, the key consideraton is the
following: all the moves the machine makes are deterministic, except for
move $\nmove{5}$, since there might be many applicable productions at that
point. If the grammar is well-formed, the following lemma is easily
seen to hold:

\begin{lemma}[Determinacy]
\label{lemma:determinacy}
Given a state $s$ of \genmachine{\cG} which allows a move $\nmove{5}$ to a
state $s'$. There exists unique states and moves
\[ s \nmove{5} s' \nmove{m_{1}} s_{1} \nmove{m_{2}} \cdots
   \nmove{m_{k}} s_{k} \]
such that $m_{1},\ldots,m_{k}\not= 5$ and state $s_{k}$ allows either no
moves or a move $\nmove{5}$.
\end{lemma}

We define a reduction relation $\cmove$ between states of
\genmachine{\cG} that allow either a move $\nmove{5}$ or no move at all:
in the statement of the above lemma, if $s \nmove{5} s'$ via
production $P$, we say that $s \cmove s_k$ via production $P$. This is
well-defined (by the above lemma) and can be seen as a collapse of
the $\move$ transitions. The following result is a reformulation of
lemma \ref{lemma:determinacy}:

\begin{corollary}
\label{corollary:determinacy}
Given $s$ a state of \genmachine{\cG}. If $s\cmove s_1$ via production
$P$ and $s\cmove s_2$ via production $P$, then $s_1=s_2$.
\end{corollary}

Let ${\cal T}$ be the transition graph of $\cmove$ generated by the
machine state \tapes{\nada}{}{}{S}{}{}{\nada}. We now show that ${\cal
L}$ is isomorphic to ${\cal T}$. Let us first define 
a mapping between strings of elements of $\cV$ and states of \genmachine{\cG}.
This function will be the isomorphism we are looking for. 

\begin{definition}
Given a grammar $\cG=(\cV,\cV_{i},\cV_{t},\cP)$, and $\sigma$ a string
of elements of 
$\cV$. Suppose $\sigma$ is of the form
t$_{1}\ldots$t$_{p}$n$_{1}\ldots$n$_{q}$P$_{1}$P$_{2}$m$_{1}\ldots$m$_{r}$,
where t$_{1},\ldots,$t$_{p}$ are prefixing
terminal symbols,
n$_{1},\ldots,$n$_{q}$,P$_{1}$,P$_{2}$,m$_{1},\ldots,$m$_{r}$ are 
nonterminal symbols and the leftmost applicable production of $\cP$ to
$\sigma$, if any, is of the form \production{P$_{1}$P$_{2}$}{...}
(P$_{1}$ might be empty). Define the function $F$ by
\[ F(\sigma) = \tapes{\nada}{$\mbox{m}_{1}\cdots\mbox{m}_{r}$}{$\mbox{t}_{1}\cdots\mbox{t}_{p}$}{$\mbox{P}_{2}$}{}{$\mbox{n}_{1}\cdots\mbox{n}_{q}$}{$\mbox{P}_{1}$} \]
or (if no production is applicable to $\sigma$)
\[ F(\sigma) = \tapes{\nada}{}{$\sigma$}{\nada}{}{}{\nada} \]
The symbols P$_{1}$ (if any) and P$_{2}$ are said to be in \emph{application
position}.
\end{definition}

\begin{lemma}
\label{lemma:1}
$F$ is injective.
\end{lemma}
\begin{proof}
Given $\sigma, \sigma' \in {\cal L}$. Assume
$F(\sigma)=F(\sigma')$. Then $\sigma=$t$_{1}\ldots$t$_{p}\sigma_{1}$
and $\sigma'=$t$_{1}\ldots$t$_{p}\sigma_{1}^{\prime}$, with
$\sigma_{1},\sigma_{1}^{\prime}$ strings of nonterminals. 
if no symbols are in application position, then by the definition of $F$
both $\sigma,\sigma'$ are strings of terminals, and by the above
$\sigma=\sigma'$. If P$_{1}$ and P$_{2}$ are in application position
(P$_{1}$ might be empty), then $\sigma_{1} =
\sigma_{2}$P$_{1}$P$_{2}\sigma_{3}$ 
and
$\sigma_{1}^{\prime}=\sigma_{2}^{\prime}$P$_{1}$P$_{2}\sigma_{3}^{\prime}$
and again by the definition of $F$, $\sigma_{2}=\sigma_{2}^{\prime}$ and
$\sigma_{3}=\sigma_{3}^{\prime}$. Thus $\sigma=\sigma'$ and $F$ is
injective. 
\end{proof}

\begin{lemma}
\label{lemma:2}
Given $\sigma,\sigma'\in{\cal L}$, then
\reduction{\sigma}{\sigma'} implies
$F(\sigma)\cmove F(\sigma')$.
\end{lemma}
\begin{proof}
Given $\sigma,\sigma'\in{\cal L}$. Assume $\sigma$ is of the
form $\tau$A$_{1}\cdots$A$_{n}$. Four cases arise, depending on the
form of the production applicable to $\sigma$ (there must be one).
\begin{enumerate}
\item \production{A$_{1}$}{t} with t a terminal symbol, and
  $\sigma'$ is of the form 
  \[ \tau\mbox{tA}_{2}\cdots\mbox{A}_{n} \]
\item \production{A$_{1}$A$_{2}$}{t} with t a terminal symbol, and
  $\sigma'$ is of the form 
  \[ \tau\mbox{tA}_{3}\cdots\mbox{A}_{n} \]
\item \production{A$_{k}$}{$\Gamma$} for some $k$, and
  $\sigma'$ is of the form
  \[ \tau\mbox{A}_{1}\cdots\mbox{A}_{k-1}\Gamma
     \mbox{A}_{k+1}\cdots\mbox{A}_{n} \] 
\item \production{A$_{k}$A$_{k+1}$}{$\Gamma$} for some $k$, and
$\sigma'$ is of the form 
\[ \tau\mbox{A}_{1}\cdots\mbox{A}_{k-1}\Gamma
   \mbox{A}_{k+2}\cdots\mbox{A}_{n} \]
\end{enumerate}
It is straightforward to show that in all those cases,
$F(\sigma)\cmove F(\sigma')$.  
\end{proof}

\begin{lemma}
\label{lemma:3}
Given $\sigma,\sigma'\in{\cal L}$, then
$F(\sigma) \cmove F(\sigma')$ implies
\reduction{\sigma}{\sigma'}.
\end{lemma}
\begin{proof}
Assume $F(\sigma)\cmove F(\sigma')$ via production 
\production{$\Gamma$}{$\Delta$}. By definition of $F$,
\production{$\Gamma$}{$\Delta$} is leftmost applicable to $\sigma$. Let
\reduction{\sigma}{\sigma^{\prime\prime}} via production
\production{$\Gamma$}{$\Delta$}. By lemma \ref{lemma:2},
$F(\sigma)\cmove F(\sigma^{\prime\prime})$ via production
\production{$\Gamma$}{$\Delta$}. By corollary \ref{corollary:determinacy},
$F(\sigma')=F(\sigma^{\prime\prime})$, and by lemma \ref{lemma:1},
$\sigma'=\sigma^{\prime\prime}$ and thus
\reduction{\sigma}{\sigma'}.
\end{proof}

\begin{lemma}
\label{lemma:4}
$F({\cal L}) = {\cal T}$.
\end{lemma}
\begin{proof}
We show by induction on $i$ that $\forall i~F(lay_{i}({\cal L})) =
lay_{i}({\cal T})$, which clearly implies the statement of the lemma.

The base case of the induction is trivial, since
$F(\mbox{S})=\tapes{\nada}{}{}{S}{}{}{\nada}$.

For the induction step, we first show $F(lay_{i+1}({\cal L}))\subset
lay_{i+1}({\cal T})$. Given $\sigma\in lay_{i+1}({\cal L})$. Thus,
there exists a $\sigma'\in lay_{i}({\cal L})$ such that
\reduction{\sigma'}{\sigma}. By the induction hypothesis,
$F(\sigma')\subset lay_{i}({\cal T})$. By lemma \ref{lemma:2},
$F(\sigma')\cmove F(\sigma)$, and by definition of transition
graph ${\cal T}$, $F(\sigma)\in lay_{i+1}({\cal T})$. 

We next show $lay_{i+1}({\cal T})\subset F(lay_{i+1}({\cal L}))$. Let
$s\in lay_{i+1}({\cal T})$. Thus there exists a $s'\in
lay_{i}({\cal T})$ with $s'\cmove s$ via production
\production{$\Gamma$}{$\Delta$}. By the induction hypothesis, there
exists a $\sigma'\in lay_{i}({\cal L})$ such that
$F(\sigma')=s'$. Let $\sigma$ be the application
of \production{$\Gamma$}{$\Delta$} to $\sigma'$. By lemma
\ref{lemma:2}, $F(\sigma')\cmove F(\sigma)$, and thus
$s'\cmove F(\sigma)$. By corollary
\ref{corollary:determinacy}, $F(\sigma)=s$ and thus $s\in
F(lay_{i+1}({\cal L}))$. This completes the induction and the proof. 
\end{proof}

\begin{lemma}
\label{lemma:5}
$F$ is an isomorphism of graphs from ${\cal L}$ to ${\cal T}$.
\end{lemma}
\begin{proof}
By lemmas \ref{lemma:1} and \ref{lemma:4}, $F$ is a bijective function
from ${\cal L}$ to ${\cal T}$. By lemmas \ref{lemma:2} and
\ref{lemma:3}, $F$ is a transition graph isomorphism.
\end{proof}

This isomorphism implies the following result for the generative
version of the production machine for a given grammar $\cG$.

\begin{proposition}
Given a grammar $\cG$, a sentence $\sigma$ is producible by
\genmachine{\cG} if and only if $\sigma$ is in the leftmost language
of $\cG$. 
\end{proposition}
\begin{proof}
($\Rightarrow$) Given $\sigma=$t$_{1}\cdots$t$_{n}$ a string in the
leftmost language of $\cG$. Thus there exists a chain in ${\cal L}$ from
S, the initial symbol of $\cG$, to $\sigma$ representing the leftmost
reductions derivation of $\sigma$. By the isomorphism of lemma
\ref{lemma:5}, there exists a chain in ${\cal T}$ 
\[ \tapes{\nada}{}{}{S}{}{}{\nada} \cmove \cdots \cmove
   \tapes{\nada}{}{$\sigma$}{\nada}{}{}{\nada} \]
Since 
\[ \tapes{S}{}{}{\nada}{}{}{\nada} \nmove{1} 
   \tapes{\nada}{}{}{S}{}{}{\nada} \]
and extending (uniquely, by lemma \ref{lemma:determinacy}) the $\cmove$
transitions, we get a sequence of machine moves
\[ \tapes{S}{}{}{\nada}{}{}{\nada} \nmove{1} \cdots \nmove{6}
   \tapes{\nada}{}{$\sigma$}{\nada}{}{}{\nada} \]
and thus $\sigma$ is producible by \genmachine{\cG}.

\vspace{.1in}
($\Leftarrow$) Given $\sigma=$t$_{1}\cdots$t$_{n}$ a string producible
by \genmachine{G}. There exists machine moves 
\[ \tapes{S}{}{}{\nada}{}{}{\nada} \nmove{1}  
   \tapes{\nada}{}{}{S}{}{}{\nada} \nmove{5} \cdots \nmove{6}
   \tapes{\nada}{}{$\sigma$}{\nada}{}{}{\nada} 
\]

Starting from \tapes{\nada}{}{}{S}{}{}{\nada} and collapsing the
$\move$ transitions into $\cmove$ transitions, we get a chain in ${\cal
T}$. By the isomorphism of lemma \ref{lemma:5}, we get a chain in
${\cal L}$
\[ \mbox{S} \reduction{}{} \cdots \reduction{}{} \sigma \]
and thus $\sigma$ is in the leftmost language of $\cG$.
\end{proof}

\section{Recognition}

Fundamentally, the recognitive machine \recmachine{\cG} is similar to
the generative one: it defines essentially the same moves (except that
the move produceing terminals is replaced by a move that accept the
next symbol from the input/output tape), and it uses the dual of the
grammar under consideration. 

One may again derive an isomorphism in the manner described in the
previous section, connecting the moves of the recognitive machine to
the leftmost reduction relation defined on the dual of the
grammar. One needs to extend the definition of transition graphs to
use strings of terminals as the initial set. The extension is fairly
trivial, and is left as an exercise.

The language generated by \genmachine{\cG} is the leftmost language of
$\cG$, the one obtained by allowing only leftmost
reductions. Correspondingly, the language recognized by
\recmachine{\cG} is a dual to the leftmost language, characterized as
those sentences that can be recognized via leftmost reductions in the
dual grammar.

It is clear that the recognized language is a subset of the full
language of the grammar. The following grammar shows that the
recognized language is in general a proper subset of the full
language, and need not be equal to the generated language: 
\begin{eqnarray*}
\mbox{S} & \productionsym & \mbox{AG} \\
\mbox{F} & \productionsym & \mbox{C} \\
\mbox{G} & \productionsym & \mbox{BC} \\
\mbox{E} & \productionsym & \mbox{AB} \\
\mbox{BC} & \productionsym & \mbox{z} \\
\mbox{A} & \productionsym & \mbox{x}
\end{eqnarray*}
The full language generated by this grammar is $\{ \mbox{xz} \}$.  
The leftmost language of this grammar is also $\{ \mbox{xz}
\}$. However, trying to recognize the string \mbox{xz} via
leftmost reductions in the dual grammar leads to a unique derivation 
\[ \mbox{xz} \productionsym \mbox{Az} \productionsym \mbox{ABC}
\productionsym \mbox{EC} \productionsym \mbox{EF} \] and thus the
string is not recognized by the machine.

\section{Conclusion}

We provide in this paper an analysis of the production machines
described by Lambek in \cite{Lambek93,Lambek93a}. We determine the
subset of the full language of a grammar that is both generated and
recognized by the machines. The generated language corresponds to the
subset of the full language one obtains by applying leftmost
reductions, and is in general a proper subset of the full 
language. Conversely, the recognized language corresponds to the
subset of the full language one obtains by applying leftmost
reductions in the \emph{dual} grammar, and is also in general a proper
subset of the full language. Moreover, the generated and recognized
language need not agree.

The generative version of production machines can in fact be reguarded
as implementing a generalized version of a Markov algorithm
\cite{Markov60,Salomaa90}. A Markov algorithm on a production grammar $\cG$
consists of repeatedly applying a leftmost applicable production to a
string, and if more than one production is leftmost applicable, the
first production (given an ordering of the productions) is applied. As
such, the algorithm is fully deterministic. In contrast, while a
generative production machine also applies leftmost applicable
productions, the choice of which production to apply if more than one
is applicable is non-deterministic.

Let us mention a possible extension of the description of the
production machines that would allow for the generation and
recognition of the full language. Recognition is the easiest to
extend: when the machine verifies all the possible choices of
production in parallel when a move $\nmove{5}$ is applicable,
one adds the parallel choice of \emph{not} applying any
production, and passing on to the next possible move of the
machine. One can extend generation in the same way, by adding a
nondeterministic choice of not applying a move $\nmove{5}$ when it is
possible to do so. This extension has a caveat: generation may fail to
produce a sentence.

An important class of grammars do not satisfy the criteria set forth
for generation and recognition via production grammars: translation
grammars, which take strings of initial symbols as initial states. For
example, the initial symbols could be words of English, and terminal
symbols words in French, and the grammar would translate English into
French. The production machines presented in this paper can be
modified easily to handle such grammars.

\section*{Acknowledgments}

Thanks to Jim Lambek for many helpful discussions and support during
this research.

\end{document}